\documentclass[aps,twocolumn,showpacs,prl,amsfonts]{revtex4}
\usepackage{epsfig,amsmath}
\usepackage{bm}
\newcommand{\Ref}[1]{(\ref{#1})}
\newcommand{\REF}[1]{Eq.~(\ref{#1})}
\newcommand{\BE}{\begin{equation}}    
\newcommand{\EE}{\end{equation}}
\newcommand{\EEA}{\end{eqnarray}} 
\newcommand{\EEa}{\end{eqnarray*}}   
\def\nn {\nonumber}  
\newcommand{\B}[1]{{\bm{#1}}}
\newcommand{\C}[1]{{\mathcal{#1}}}    
 \newcommand{\ve}{\varepsilon} 
\newcommand{\p}{\partial}           

\def\<{\left \langle} \def\>{\right\rangle}
\def\Re{{\C R}\mkern-3.1mu e}  

\renewcommand{\sb}[1]{_{\text {#1}}}  

\begin{document}
\title{ Energy spectra of developed superfluid turbulence}

\author{Victor S. L'vov} \email{Victor.Lvov@Weizmann.ac.il}
\homepage{http://lvov.weizmann.ac.il} \affiliation{ Department of
Chemical Physics, The Weizmann Institute of Science, Rehovot
76100, Israel}
\author{Sergey V. Nazarenko}\email{snazar@maths.warwick.ac.uk}
\homepage{http://www.maths.warwick.ac.uk/~snazar} 
\affiliation{University of Warwick, Mathematics Institute,
Coventry, CV4 7AL}
\author{Grigory E. Volovik},
\email{Volovik@boojum.hut.fi} \affiliation{Low Temperature
Laboratory,
Helsinki University of Technology, P.O.Box 2200, FIN-02015 HUT, Finland\\
L.D. Landau Institute for Theoretical Physics, Kosygin Str. 2.
117940 Moscow, Russia}
\begin{abstract}
Turbulence spectra in superfluids are modified by the
nonlinear energy dissipation caused by the mutual friction between
quantized vortices and the normal component of the liquid.
 We have found a new state of fully  developed
turbulence which occurs in some range of
two Reynolds parameters characterizing the superfluid flow.
This state displays both the Kolmogorov-Obukhov
$\frac53$-scaling law $E_k \propto k^{-5/3}$ and a new
``$3$-scaling law" $E_k \propto k^{-3}$, each in a well-separated
range of $k$.
\end{abstract}
\date{\today}
\maketitle
Superfluid
consists of mutually penetrating
components --
 viscous normal component and one or several
frictionless superfluid components.
This explains why
 different types of
turbulent motion are possible depending on whether the normal and
the superfluid components move together or separately.
Here, we are
interested in the most simple case when the dynamics of the normal
component can be neglected. This occurs, for example, in the
superfluid phases of $^3$He where the normal component is so
viscous that it is practically clamped to the container walls. The
role of the normal component in this case is to provide the
preferred heat-bath reference frame, where the normal
component,
and thus the heat bath,
 are at rest. Dissipation takes place
when the vortices move with respect to this reference frame.
Turbulence in such a superfluid component with the normal component
at rest will be called here superfluid turbulence.

Recent experiments in $^3$He-B \cite{Finne} demonstrated that the
fate of
a few vortices injected into a rapidly moving superfluid
depends on a dimensionless intrinsic
temperature-dependent parameter
$q$
 rather than on the flow velocity. At $q \sim 1$, a
rather sharp transition is observed between
 laminar evolution
of the injected vortices and
a turbulent many-vortex
state of the whole superfluid. This
adds a new twist to the
general theory of turbulence in superfluids developed by Vinen
\cite{Vinen,VinenNiemela} and others. Attempts to modify the
theory in order to incorporate the new phenomenon, have been made
in Refs. \cite{Kopnin}, \cite{Volovik2003} and \cite{VinenNew}.

In this Letter we describe how
the celebrated Kolmogorov-Obukhov
$\frac53$-law for the turbulent energy spectrum in normal fluid,
$E_k\propto k^{-5/3}$,  gets modified  in the superfluid turbulence,
giving rise the much steeper decrease, $E_k\propto k^{-3}$.

As a starting point we  utilize a coarse-grained hydrodynamic
equation for the superfluid dynamics with distributed
vortices.  In this equation the parameter $q$ characterizes  the
friction force between the superfluid and the normal components of the
liquid, which is mediated by quantized vortices. According to this
equation, turbulence develops only if the friction is relatively
small compared to the inertial term, i.e. when $q < 1$.
Here, we will study the case of  developed turbulence which must occur at
$q\ll 1$.

An important feature of  superfluid turbulence is that the
vorticity of the superfluid component is quantized in terms of the
elementary circulation quantum $\kappa$ (in $^3$He-B,
$\kappa=\pi\hbar/m$ where $m$ is the mass of $^3$He atom). Thus,
superfluid turbulence is a chaotic motion of well-determined and
well-separated vortex filaments \cite{VinenNiemela}. Using this as
starting point we can simulate the main ingredients of classical
turbulence -- the chaotic dynamics of the vortex degrees of
freedom of the liquid. However, to make the analogy useful for
classical turbulence one must choose the regime described by
equations of  the hydrodynamic type valid at  length-scales above
the inter-vortex distance, the latter being a microscopic
cut-off similar to the inter-atomic distance in conventional
hydrodynamics. The coarse-grained hydrodynamic equation for the
superfluid component  is obtained from the Euler equation for the
superfluid velocity  ${\bf v}\equiv {\bf v}_{\rm s}$  after
averaging over the vortex lines (see review \cite{Sonin}):
\begin{equation}
\frac{\p \B v}{\p t}+(\B v \cdot \B \nabla) \B v+ \B \nabla \mu=
\B D\,, \label{SuperfluidHydrodynamics1}
\end{equation}
where   $\mu$ is the chemical potential and $\B D$ describes the
mutual friction:
\begin{equation}
\B D= -\alpha'({\bf v} -{\bf v}_{\rm n})\times \B \omega+
\alpha~\hat{\B \omega}\times[\B \omega \times({\bf v} -{\bf
v}_{\rm n}) ] ~. \label{SuperfluidHydrodynamics2}
\end{equation}
Here $\B \omega=\nabla\times {\bf v}$ is the superfluid vorticity;
$\hat{\B \omega}=\B \omega/\omega$; ${\bf v}_{\rm n}$  is the
velocity of the normal component (which is fixed); $\alpha'$ and
$\alpha$ are dimensionless parameters describing the mutual
friction between superfluid and normal components of the liquid
mediated by quantized vortices which transfer momenta from the
superfluid to the normal subsytem. For the flow with vortices
locally aligned with each other these parameters enter the
reactive and dissipative forces acting on a vortex line as it
moves with respect to the normal component. For vortices in
fermionic systems (superfluid $^3$He and superconductors) such
forces acting on a vortex were calculated by Kopnin \cite{KopninBook},
and they were measured in $^3$He-B over a broad temperature range
by Bevan et al. \cite{Bevan}.  Here we consider $\alpha'$ and
$\alpha$ as phenomenological parameters, assuming the general case
where quantized vortices are not aligned locally and thus the bare
parameters are renormalized.

Further we shall work in the reference frame where ${\bf v}_{\rm
n}=0$. In this frame, the nondissipative  first term in
Eq.~(\ref{SuperfluidHydrodynamics2}) renormalizes the inertial
term ${\bf v} \times \B \omega$ in the left hand side (LHS)  of
Eq.~(\ref{SuperfluidHydrodynamics1}) by the factor $1-\alpha'$.
The role of the Reynolds number in this hydrodynamics, i.e.
relative magnitude of the two non-linear terms, the inertial and
friction ones, is played by the velocity independent ratio of
dimensionless parameters, $\Re =(1-\alpha')/\alpha$, which was
denoted in Ref. \cite{Finne} as $\Re =1/q$. The role of the
parameter $q$ as the inverse Reynolds number was demonstrated  in
experiments of Ref. \cite{Finne}, where it was shown that the
turbulence develops only below some critical value of $q$ (i.e. at
$q<q_c \sim 1$).

Here we are interested in the region of large  Reynolds numbers,
$\Re \gg 1$ ($q\ll 1$), where the inertial term is strongly
dominating. In this region one expects well developed
turbulence characterised by a Richardson-Kolmogorov-type
cascade which is modified  due to the non-linear dissipation. In
$^3$He-B, the range $q\ll 1$ occurs at low temperatures,
where  $\alpha'\ll \alpha$,  $q\approx \alpha$. Then the mutual
friction term in Eq.~(\ref{SuperfluidHydrodynamics2})  can be
written as
\begin{equation}\label{V^{2}} 
\B D =q\,  \B \omega\times [\B \omega \times \B v]/| \omega |\,,
\end{equation} 
and  we finally arrived at the hydrodynamic equation
(\ref{SuperfluidHydrodynamics1}) whose LHS is the usual Euler
equation  with the nonlinear term which is responsible for the
energy cascade in the developed hydrodynamic turbulence, while the
RHS contains the nonlinear dissipation term given be
Eq.~(\ref{V^{2}}). We will describe this cascade in the simplest
possible manner, using the differential form~\cite{Kov,Leith} of
the energy transfer term in the energy budget equation (the more
complicated  version with second derivative
\cite{ConnaughtonNazarenko}  was  used for superfluid turbulence
by Vinen \cite{VinenNew}):
\begin{equation}\label{V3} 
\frac{\p E_k}{\p t}= - \C D_k - \frac{\p \ve_k}{\p k}\ .
\end{equation}
Here $E_k$ is the one-dimensional density of the turbulent kinetic
energy in the $k$-space,  defined such the the total energy
density (in the physical space) $E$ is given by
\begin{equation}\label{V4} 
E\equiv \frac12\, \< |\B v|^2 \>=\int dk\, E_k\ .
\end{equation}
Dissipation of energy on scale $k$ is described by the $\C D_k$
term in the right hand side (RHS) of \REF{V3} which will be
clarified later. The idea of \cite{Kov} is to relate $E_k$ and
$\ve_k$ in \REF{V3} in the spirit of Kolmogorov 1941 (K41) dimensional
reasoning:
\begin{equation}\label{V5} 
E_k=C \ve^{2/3}_k k^{-5/3}\ .
\end{equation}
Here  $C\simeq 1$ is the Kolmogorov dimensionless constant. In the
absence of dissipation,  \REF{V3} immediately produces the
stationary solution $\ve_k=\ve$ with constant energy flux
$\ve$ in the inertial interval of scales. Then  \REF{V5} turns
into  the Kolmogorov-Obukhov $\frac53$-law for $E_k$:
\begin{equation}\label{V6} 
E_k=C \ve^{2/3} k^{-5/3}\ .
\end{equation}

The goal of this Letter is to describe possible modification of the
scaling exponents in $E_k$  due to different, than in the
Navier-Stokes equation, form \Ref{V^{2}} of  the dissipation term.
The energy balance equation in the differential approximation \Ref{V3}
is just an adequate tool for this study: it is as simple as
possible, but not more. More accurate integral representation for
the energy transfer term  gives exactly the same results for the
scaling exponents, because they  arise from the power counting.
Clearly, approximation \Ref{V3} does not accurately control
numerical prefactors and the exact functional form of the possible
crossover region, but these are not important for the questions we
aim to study here.

Within the same level of accuracy, we can simplify the vectorial
structure of the dissipation term $\B D$ and average \REF{V^{2}}
over the directions of the vorticity $\B \omega$ (at fixed direction
of $\B v$)
\begin{equation}\label{V7} 
\B D \Rightarrow \< \B D \>  _{\B \omega / |\omega|}=
  -\frac23 \,q\, |\omega| \B v\Rightarrow
- \,q\, |\omega| \B v \ .
\end{equation}
Again, we are not bothered by the numbers and therefore skipped
for simplicity  factor $2/3$ in the last of Eqs. \Ref{V7}. Notice
that vorticity in hydrodynamic turbulence is usually
dominated by  the $k$-eddies (i.e. motions of scale $\sim 1/k$) with
the largest characteristic wave-vector $k\sb{max}$. These eddies
have the smallest turnover  time $\tau\sb{min}$ that is of the order of
their decorrelation time. One can show that the main contribution
to the velocity (not vorticity) in the equation for the
dissipation of the $k$-eddies, $\C D_k$, with intermediate
wave-vectors $k$,  $1/R <k \ll k\sb{max}$,   is dominated by
the $k'$-eddies with $k'\sim k$. Because the turnover  time of these
eddies $\tau_{k'}\gg  \tau\sb{min}$, we can think of $|\omega|$ in
\REF{V7} on time intervals of interest ($\tau\sb{min} \ll  \tau\ll
\tau_{k'}$) as self-averaging
 quantity because it is almost uncorrelated
with the velocity $\B v$  which can be treated as  dynamical variable.
In this study, this allows us to neglect in \REF{V7} the fluctuating
part of $|\omega|$ and  to replace $|\omega|$ by its mean value.
In this approximation \REF{V7} takes very simple form:
\begin{equation}\label{V8} 
\B D  = - \Gamma\, \B v \,, \quad  \Gamma \, \equiv  q\, \omega_0
\,, \quad \omega_0  \equiv  \< |\omega|\>\ .
\end{equation}

From  \REF{V8} one easily finds that
\begin{equation}\label{V9} 
\C  D_k  = \Gamma\, E_k\,,
\end{equation}
and the balance equation \Ref{V3}  in the steady state
finally takes the form:
\begin{equation}\label{V10} 
{\partial \ve _k \over \partial k} = -2 \, \Gamma \ve _k ^{2/3}
k^{-5/3}\,,
\end{equation}
in which for the simplicity we put $C=1$, because in our simple
approach we are not controlling numbers of the order of unity.

Let us analyze the  solutions of \REF{V10} that  arise in presence
of a fixed energy influx $\epsilon_k= \epsilon_+$ at $k=1/R$ into
the turbulent system.  Hereafter  $R$ is the outer scale of
turbulence that of the order of the radius of the cryostat.    In
terms of dimensionless variables, $p = kR, \; f_p = \epsilon_p /
\epsilon_+, \; \gamma = \Gamma V /R $ and $V=(\epsilon_+R)^{1/3}$,
we have
\begin{equation}\label{Eq}
{\partial f_p \over \partial p} = -2 \gamma f_p^{2/3} p^{-5/3}.
\end{equation}
The solution to this with the boundary condition $f|_{p=1} = 1$ is
\begin{equation}\label{V13}
f_p^{1/3} = \gamma p^{-2/3} + 1 - \gamma,
\end{equation}
or, in terms of the dimensional energy spectrum,
\begin{equation}
E_k = {V^{2} \over k (kR)^{2/3}} \left[1 + {\gamma \over
(kR)^{2/3}} - \gamma \right]^2. \label{E}
\end{equation}
Then, expressing  the mean vorticity  through  this energy
spectrum,  one obtains  the closed equation for the parameter
$\gamma$:\begin{equation} \Gamma= \gamma\, R/V
=q\omega_0(\gamma)~,\label{gammaEq}
\end{equation}
which manifests the existence of several different  regimes of
turbulence. Consider first the case when the resulting $\gamma <
1$ and Eq. (\ref{E}) can be rewritten as
\begin{equation}
E_k = {V^{2} \gamma^{2} \over R^{2} k^{5/3}} \left[{1 \over
k^{2/3}} + {1 \over k\sb{cr}^{2/3}} \right]^2 \, , \label{E1}
\end{equation}
where
\begin{equation}
k\sb{cr}\equiv {\gamma^{3/2} \over (1-\gamma)^{3/2}R} \label{kcr}
\end{equation}
is a crossover wavenumber separating two different scaling ranges.
For $ k \gg k\sb{cr}$ we have the K41 scaling in which the
dissipation is negligible and the energy flux is approximately
constant,
\begin{equation}
E_k = {V^{2}  (1-\gamma)^{2} \over R^{2/3}k^{5/3}}  \ .
\label{k41}
\end{equation}
This equation can be rewritten in the traditional form~\Ref{V6}
\begin{equation}\label{V19} 
E_k\simeq   \ve^{2/3}_\infty  k^{-5/3}\,,
\end{equation}
in which the energy flux $\ve_\infty$ for  $ k > k\sb{cr}$ due to
the mutual friction  can be much smaller than the energy influx,
$\ve_+$, into the turbulent system:
\begin{equation}\label{V^{2}0} 
\ve_\infty = \ve_+  (1-\gamma)^{3}< \ve_+\ .
\end{equation}

In order to see how $\ve_k$ decreases toward large $k$,
approaching $\ve_\infty$ at $k\sim k\sb{cr}$ consider region $k
\ll k\sb{cr}$. In this region Eqs.~\Ref{V13}, \Ref{E} yield:
\begin{equation}
\ve_k=\frac{\ve_+}{ (k R)^{2}}\,, \quad E_k = {V^{2}  \gamma^{2}
\over R^{2}\, k^{3}}\ . \label{new}
\end{equation}
Thus the rate of the energy  dissipation, being proportional to
$E_k$, decreases toward large $k$ and becomes insignificant at  $
k \gg k\sb{cr}$. In this region $\ve_k\simeq \ve_\infty$ and one
has the K41 scaling~\Ref{V19}.

The K41 scaling ends by a cutoff at a``microscopic" scale $1/k_*$
at which circulation in the $k_*$-eddy reaches the circulation
quantum $\kappa$:
\begin{equation}
\kappa \sim {v_* \over k_*} = {R V (1- \gamma) \over
(k_*R)^{4/3}}, \label{quant}
\end{equation}
i.e.
\begin{equation}
(k_*R)^{4/3} \sim {\cal N}^2  (1-\gamma)\,, \label{kstar}
\end{equation}
where ${\cal N}^2 =R V/\kappa$ is the  ``quantum Reynolds number''
\cite{Finne}. Parameter $\C N$ can be considered as the ratio of
$R$ to the mean inter-vortex distance. Clearly, with the classical
approach to the problem we can consider only the limit $\C N \gg
1$.

Now we can clarify the equation of self-consistency
(\ref{gammaEq}) for $\gamma$. Estimating  $\omega_0$ as follows
$$
\omega^2_0=\<|\omega|\>^2\simeq \<|\omega|^2\>\simeq
\int\limits_{1/R}^{k_*}dk \, k^{2}\,  E_k\,,
$$
and using Eqs.~\Ref{gammaEq} and \Ref{E1} one gets
\begin{equation}
1 \simeq q^{2} \int\limits_{1/R}^{k_*}dk \, k^{1/3}\left[{1 \over
k^{2/3}} + {1 \over k\sb{cr}^{2/3}} \right]^2  \ . \label{self1}
\end{equation}
Together with Eqs.~\Ref{kcr} and \Ref{kstar}, that relate $k_*$
and $k\sb{cr}$ with $\gamma$ and $\C N$,  this equation allows to
find $\gamma$, $k_*$ and $k\sb{cr}$ in the terms of ``external
parameters" of the problem, $q$ and $\C N$. As we pointed out, the
classical regime of developed turbulence corresponds to the region
$q\ll 1$, $\C N \gg 1$.

Consider first the  case when the inner (quantum) scale of
turbulence is well separated from the crossover scale: $k_*\gg
k\sb{cr}$. Then the main contribution to the integral in
\REF{self1} comes from the region $k\gg k\sb{cr}$, when the second
term in the integral dominates. Therefore \REF{self1} gives the
relationship
\begin{equation}
k\sb{cr} \simeq q^{3/2}  k_*\,, \label{self2}
\end{equation}
which together with Eqs.~\Ref{kcr} and \Ref{kstar} yields Eq. for
$\gamma$:
\begin{equation}
q\, \C N \simeq \gamma/(1-\gamma)^{3/2}\ . \label{gammaEq1}
\end{equation}
For $q\C N \ll 1 $ the solution  is $\gamma\simeq q \C N < 1$ that
gives $R k\sb{cr}\simeq (q\, \C N)^{3/2}< 1$ and:
\begin{eqnarray}\nn
E_k &\simeq&  {V^{2} \over k (kR)^{2/3}} \left[1 - 2\, q\C N
\left(1-\frac1{(k R)^{2/3}}\right)\right]\,,\\
R k_*& \simeq&  \C N^{3/2}\gg 1\,, \qquad \mbox{for} \quad q\C
N\ll  1\ . \label{E4}
\end{eqnarray}%
It means that for $q\C N \ll 1 $ in the entire inertial interval,
$1/R<k<k_*$, one has usual Kolmogorov-Obukhov spectrum with the
small, of the order of $q \C N$, negative corrections. In other
words, at $q\C N \ll 1 $ the mutual friction has a negligible
effect on the statistics of turbulence.

The situation is different in the region $q\C N \gg 1$. In this
case the solution to \REF{gammaEq1} is $1-\gamma\simeq (q\C
N)^{-2/3}\ll 1$ and instead of \REF{E4} one has:
\begin{eqnarray}\nn 
E_k&\simeq& {V^{2} \over R^{2} k^{5/3}} \left[{1 \over k^{2/3}} +
{1 \over k\sb{cr}^{2/3}} \right]^2 \, , \qquad \mbox{for}\quad q\C
N \gg 1\,,\\
k\sb{cr} R &\simeq&   q  {\cal N}\,, \quad k_*R \simeq  {\cal N}
/q^{1/2}\gg k\sb{cr}R\ . \label{E3}
\end{eqnarray}%
One sees  that the pumping  and the crossover scales are well
separated if $q{\cal N}\gg 1$. Under this condition, the quantum
cutoff scale is also well separated.

 Up to now we assumed that the second  term in $[\dots]$ in the
integral~\Ref{self1} is dominant.  In the opposite case instead of
Eqs.~\Ref{self2} and \Ref{gammaEq1} one has:
\begin{equation} \label{self3}
k_*R\simeq \exp \left( \frac1 {q^{2}}\right)\,, \quad
k\sb{cr}R\simeq \C N^3\exp \left( - \frac2 {q^{2}}\right)\ .
\end{equation}
Taking into account that  we are considering solutions in which
the first
  term in $[\dots]$ in the integral~\Ref{self1} is dominant
we have to take $k\sb{cr}>k_*$ which gives $\C N>\exp(1/q^2)$.
Therefore the range of parameters where the two-cascade
regime~\Ref{E3} occurs is
\begin{equation} 1/q < {\cal N} < \exp(1/ q^{2})~.
\label{Conditions1}
\end{equation}
An important feature of this solution is that both the energy
spectrum $E_k$ and the spectrum of the flow dissipation $\ve_k$
are concentrated at the largest length scale $R$, whereas the
dissipation is mediated by the vorticity $\omega_0$ concentrated
at the smallest (microscopic) scale $1/k_*$. Therefore the energy
balance between the Kolmogorov cascade and the energy dissipation
must occur already for the largest eddies. This gives the
condition \cite{Volovik2003}
\begin{equation}
{V^{3}\over R}=\Gamma V^{2}~. \label{BalanceCondition}
\end{equation}
This means that in this turbulent state the mean vorticity is
$\omega_0=\Gamma/q=U/qR$. If $q{\cal N}\gg 1$ one has $\gamma$
close to 1   and thus our double-cascade solution~\Ref{E3}
satisfies the large-scale balance (\ref{BalanceCondition}).

At $q{\cal N}\sim 1$ one has $k\sb{cr}=1/R$, i.e.  the region of
the $k^{-3}$ spectrum shrinks. At $q{\cal N}\ll 1$ the parameter
$\gamma$ deviates from unity, $\gamma\simeq q\C N$. Here two
scenarios  are possible. In the first one  we have
solution~\Ref{E4} in which the mutual friction is unessential and
thus is unable to compensate the Kolmogorov cascade. When the
intervortex distance scale is reached the Kolmogorov energy
cascade is then transformed to the Kelvin wave cascade
\cite{VinenNiemela} for the isolated vortices. In the second
scenario suggested in Refs. \cite{Volovik2003} and
\cite{VolovikJLTP}, at $q{\cal N}\ll 1$ the turbulent state is
completely reconstructed and the so called Vinen state emerges.
This state introduced by Vinen \cite{VinenOld} and then by Schwarz
\cite{Schwarz} contains a single scale $r=\kappa/V$ and thus no
cascade.

Another interesting case to consider is $\gamma>1$ when
\REF{E} for $E_k$ can be rewritten as follows:
\begin{equation}
E_k = {V^{2} \gamma^{2} \over R^{2} k^{5/3}} \left[{1 \over
k^{2/3}} - {1 \over \tilde  k\sb{cr}^{2/3}} \right]^2\,,
\label{E5}
\end{equation}
where
\begin{equation}
\tilde k\sb{cr} = {\gamma^{3/2} \over (\gamma-1)^{3/2}R}~.
\label{kcr2}
\end{equation}
Here the Kolmogorov cascade stops already at the scale $\tilde
k\sb{cr}$, and the equation for $\gamma$ reads:
\begin{equation}
1\simeq q \sqrt{\frac32 \,\ln {\gamma \over (\gamma-1) } }  ~.
\label{gammaExotic}
\end{equation}
The solution satisfying the large-scale energy balance
$\gamma\simeq 1$ is
\begin{equation}
\gamma- 1\simeq \exp(-2/3 q^{2}) ~. \label{gammaExotic2}
\end{equation}
This solution is self-consistent and does not require the
microscopic scale cut-off if the circulation at the scale $ \tilde
k\sb{cr}$ is big enough, i.e. $ v\sb{cr}/ \tilde k\sb{cr}\gg
\kappa$. This occurs, however, at very large counterflow ${\cal
N}\gg \exp(1/q^{2})$.  However, we think that this solution   is
unstable. Indeed, let us consider a perturbation of this spectrum
in a form of cutting off its tail. This will lead to a reduction
in the dissipation rate so that the subsequent evolution will
build a a K41 constant-flux tail rather than restore the $k^{-3}$
scaling. The K41 tail will strengthen until its contribution to
the friction will restore the energy balance. Thus, the resulting
new steady state will have the K41 part, i.e. will be of the first
kind.

The summary of the different regimes (without possible Vinen
state) is shown in the Table.
%
%
\begin{table}
\begin{center}
\begin{tabular}{||c||c|c||}
\hline\hline
Intensity, $\C N$      & Crossover, $k\sb{cr} R$ & Quantum cutoff, $k_*R$  \\
\hline\hline
$\displaystyle  1<  \C N < \frac1q $  &  None, $\frac53$-scaling & $ \C N^{3/2}$ \\
$\displaystyle 1< \frac1q< \C N   < e^{1/q^2}$ & $\displaystyle {q \C N}$
& $\displaystyle\frac{ \C N}{\sqrt{q} } $ \\
$1<  e^{1/q^2} <{\C N}  $ &  None, $3$-scaling & $ \displaystyle{ e^{1/q^2}}$ \\
\hline\hline
\end{tabular}
\caption[]{Scaling range boundaries in the cases of weak,
intermediate and strong pumping.} \label{tab:ranges}
\end{center}
\end{table}

Now let us compare our results for the superfluid turbulence with
earlier works. In Refs. \cite{Volovik2003} and \cite{VolovikJLTP}
the effect  of the mutual friction on the high momentum tail was
overestimated, which led to the incorrect result for the spectrum
of dissipation at high momenta. However, some general features
suggested in Ref. \cite{Volovik2003} remain the same. In
particular, there are different regimes of the superfluid
turbulence. The transition line between two turbulent regimes,
$q\C N\simeq 1$, was correctly determined in Refs.
\cite{Volovik2003,VolovikJLTP}, as well as the vorticity in the
regime of cascade, $\omega_0=V/(qR)$. Recently Vinen
\cite{VinenNew} used a diffusion equation model which  is similar
in spirit to the model used in our Letter and, perhaps, even
better because in properly accounts not only for the cascade
states but also for the thermodynamic equilibria. However, this
equation is harder to solve analytically and Vinen used numerical
simulations. his qualitative and numerical results are consistent
with our analytical solution.

In conclusion, we discussed the spectrum of the superfluid
turbulence governed by the nonlinear energy dissipation due to the
mutual friction between the vortices and the normal component of
the liquid, which remains at rest. We found that in agreement with
Refs. \cite{Finne,Volovik2003}, the flow states are determined by
two dimensionless parameters: the velocity-independent Reynolds
number $\Re =1/q$ which separates the laminar and turbulent
states;  and the quantum velocity-dependent parameter ${\cal
N}=\sqrt{VR/\kappa}$ which contains the quantum of circulation
around the quantum vortices $\kappa$ and which determines the
transition, or crossover, between different regimes of superfluid
turbulence. In some region of the $(q,{\cal N})$ plane, we found
the turbulent state with a well defined Richardson-type cascade.
This state displays both the Kolmogorov-Obukhov $\frac53$-scaling
law $E_k \propto k^{-5/3}$ and the new ``$3$-scaling law" $E_k
\propto k^{-3}$, each in a well separated range of $k$.
Possible connection of the phase diagram of the flow states to
experimental observations is discussed in Ref. \cite{Skrbek}.

This work was done as the result of collaboration during the
workshop on ``Turbulence and vacuum instability in condensed
matter and cosmology'' organized in the framework of the ESF
Programme COSLAB. GEV thanks V.V. Lebedev and W.F. Vinen for the fruitful
criticism. The work is supported by ULTI-4 and is also supported
in part by the Russian Foundation for Fundamental Research under
grant $\#$02-02-16218 and the US-Israel Binational Science
Foundation.

\end{document}